# The Everett Interpretation: Probability [1]

Simon Saunders

The Everett interpretation of quantum mechanics is, inter alia, an interpretation of objective probability: an account of what probability really is. In this respect, it is unlike other realist interpretations of quantum theory or indeed any proposed modification to quantum mechanics (like pilot-wave theory and dynamical collapse theories); in none of these is probability itself the locus of inquiry. As for *why* the Everett interpretation is so engaged with the question of probability, it is in its nature: its starting point is the unitary, deterministic equations of quantum mechanics, and it introduces no hidden variables with values unknown.

Does it explain what objective probability is, or does it explain it away? If there are chances out there in the world, they are the *branch weights*. All who take the Everett interpretation seriously are agreed on this much: there is macroscopic branching structure to the wave-function, and there are the squared amplitudes of those branches, the branch weights. The branches are worlds – provisionally, worlds at some time. The approach offers a picture of a branching tree with us at some branch, place, and time within. But whether these weights should properly be called "chances" or "physical probabilities" is another matter. For some, even among Everett's defenders, it is a disappearance theory of chance; there *are* no physical chances; probability only lives on as implicit in the preferences of rational agents, or as a "caring measure" over branches, or in degrees of confidence when it comes to the confirmation of theories or laws; probability has no place in the physics itself.[2] The interpretation was published by Hugh Everett III in 1957 under the name "'Relative state' formulation of quantum mechanics"; he named a much longer manuscript "Wave Mechanics Without Probability."[3]

Dissent on this point among Everett's defenders is significant. If the basic category of probability is to be abolished, Everett's approach can hardly claim to be an *interpretation* of quantum mechanics: for is not the theory couched in terms of the language of probability? Critics may well conclude that their work has been done for them, but for three reasons they should think again:

First, because many of the criticisms that apply to probability in the Everett interpretation apply to every other half-way serious theory of physical probability. The difficulties may only be more vivid, more obvious, in the Everett interpretation; that may be to its credit.[4]

Second, because in philosophy of probability over the last three decades (and a mark of the influence of David Lewis's writings) a great deal of attention has been devoted to "reductive" theories of chance: theories that start with a non-chancy, "base" level of properties and relations ("Humean" properties and relations), on which probabilities, if any, are to supervene. Everett called his work "wave mechanics without probability" for good reason: it provides a non-chancy base level of categorical properties and relations, deterministically evolving in time. It fits this mold – branching and branch amplitudes don't *seem* to involve chance. This is typical of reduction: neither do wavelengths of light and relative spectral reflectancies of surfaces *seem* to involve colours. Moreover, all this philosophical time and energy has only gone to show just how difficult the chance concept is in comparison to other more successful reductive projects (of colours and other secondary qualities, causation, modality, persons, and even mind). It was with chance, famously, that Lewis feared his "Humean supervenience" project would fail. Naïve frequentism, the view that chances at $t$ are relative frequencies up to time $t$, fails for well-known reasons.[5] The proposal that they may depend on future relative frequencies too, one of Lewis's main innovations, falls prey to "undermining" (see Section 5). But the Everett interpretation of quantum mechanics provides a different supervenience basis (a branching structure) and new primitive relations (relations in phase and amplitude), and the differences appear to be decisive: armed with these, the usual criteria for reductive theories of chance can be met to perfection.

Third, among those criteria for a successful reductive theory of chance there is one that has rarely been met even in isolation: "the decision-theory link", the link between physical probability and rational belief. As distilled by Lewis and others, it is to explain, or justify, the "Principal Principle", roughly speaking the principle that if you know the chance of $E$ at $t$ is $x$, then your credence in $E$ at $t$ should be $x$. The theory of probability as it was developed by Blaise Pascal and Pierre-Simon Laplace had always included this link, for they based probability on a principle of indifference, in turn based on the symmetries of actual things. That worked well for games of chance (where dice, coins and cards have obvious symmetries), but less so for subsequent applications of probability theory to physics.[6] In the mathematical development probability theory in terms of measure theory and Boolean algebras by Emile Borel at the end of the 19th century, the link with symmetries was lost. Principles of indifference have played little role in philosophy of probability since.

In this context the observation made by David Deutsch in 1999 that certain symmetries in quantum theory force a principle of indifference is of great importance. His further demonstration that by the use of various ancillary devices, the principle in effect forces the Born rule is game-changing. The argument was substantially strengthened by David Wallace, making do with considerably weaker axioms of decision theory, supplemented instead by explicit appeal to the decoherence-based Everett interpretation. Wallace's book *The Emergent Multiverse* published in 2012 is a landmark in the foundations of both probability and quantum mechanics: nothing comparable for the decision-theory link has been achieved in any one-world theory of chance, however fanciful, let alone one based on any extant science.

It is not any old science. Quantum mechanics, by a long shot, is the most accurate, prolific, and unifying physical theory that has ever been seen – while yet, somehow, remaining the least understood. If the Everett interpretation is correct, it explains that fact as well. If in reality there is macroscopic branching satisfying the Schrödinger equation, and no hidden variables, no wonder quantum mechanics is so difficult to understand for those (the great majority) intent on a one-world interpretation. In contrast, none of the usual paradoxes of quantum mechanics pose any problem to the Everett interpretation: the measurement problem is solved, the appearance of Bell non-locality is explained,[7] and no special posits are needed, over and above the assumption that the Schrödinger equation applies to everything.

## 1 The Connection with Uncertainty

Many of the conceptual questions that arise with probability on the Everett interpretation of quantum mechanics arise equally in one-world theories, but one stands out:[8] unlike in classical statistical mechanics and hidden-variable quantum theories, and unlike in a stochastic dynamical theory, there seems to be no room in the theory for the usual connection between chance and uncertainty. Knowing the quantum state and given a unitary deterministic process (the Schrödinger equation), in the absence of any additional and "hidden" variables, when a simple quantum experiment is performed the result is a superposition of all the outcomes, with varying phases and amplitudes. If that is all that there is, it seems we know everything there is to know of any salience – but then it seems there can be no place for uncertainty; and with no uncertainty, then there can be no probability either. It is difficult even to say what credences could possibly mean: degrees of belief in what, exactly?

I offer three answers, increasingly deflationary, all consistent with one another. The first is that the model of measurement on which the argument rests is only a fragment of a realistic analysis. The branching structure of macroscopic bodies involves much more – inter alia, it requires the quantum histories formalism and the theory of quasiclassical domains (as argued in a companion paper, "The Everett interpretation: structure"[9]). In this formalism, there are superpositions of worlds understood as serial quasiclassical histories that extend to future times. The perspective is structural, a "quantum block universe", using the Heisenberg-picture. But from this perspective there is an obvious candidate for ignorance: we do not know which history is our own. Uncertainty is self-locating uncertainty, and degrees of belief are beliefs about one's location among all these histories.

A similar conclusion can be arrived at from a number of different philosophical directions, beginning from any of persons, persistence, language use, or identity. Take personhood: what are persons in terms of the Everettian reductive base? In a one-world setting a popular answer is that they are continuants, four-dimensional histories, spatially localized at each time (so spacetime worms or world tubes). The same can be taken over to branching structure (work backward, from arbitrary future times). Then given the usual attribution of speech acts to persons, ignorance in the face of branching is inevitable. If a *person* asks, at time *t*, prior to branching, what happens next *to her*, she cannot possibly know.[10]

This may seem like a cheat. Normally expressions that use indexicals like "here" or "now" or personal pronouns like "she" are tagged to places, times, persons, in a way that is causally informative. Prior to branching, how do persons that only differ with respect to future contingents presently pick themselves out? But they don't have to: the perspective is non-local in time, as is appropriate to human agency.

A second argument for uncertainty is a stripped-down version of the first. According to this, uncertainty is still bound up with self-location, but it is localized to the chance process. This will need some stage setting. Consider a simple schematic measurement process, say the measurement of the *z*-component of spin in the Stern-Gerlach experiment. Let the apparatus or observer be initially in a "ready" state $|0\rangle$, and suppose that when presented with a spin system with positive *z*-component of spin in the state $|\phi_+\rangle$ it passes to a state in which it registers spin plus, denote $|+\rangle$, and presented with a state $|\phi_-\rangle$ it passes to a state in which it registers spin minus, denote $|-\rangle$. That is, the unitary evolution $U_M$ (the subscript "$M$" is for measurement) should satisfy:

$$|\phi_+\rangle \otimes |0\rangle \xrightarrow[U_M]{} |\phi'_+\rangle \otimes |+\rangle \qquad (1a)$$

$$|\phi_-\rangle \otimes |0\rangle \xrightarrow[U_M]{} |\phi'_-\rangle \otimes |-\rangle. \qquad (1b)$$

(There is no need to assume the experiment is repeatable; hence the primes on the spin states following the measurement, which could be anything.) For an initial state

$$|\psi\rangle = c_+ |\phi_+\rangle + c_- |\phi_-\rangle \qquad (2)$$

where $c_+$ and $c_-$ are non-zero complex numbers it follows from unitarity that

$$|\psi\rangle \otimes |0\rangle \xrightarrow[U_M]{} c_+ |\phi'_+\rangle \otimes |+\rangle + c_- |\phi'_-\rangle \otimes |-\rangle. \qquad (3)$$

There are the two states at the end, in a superposition, two macroscopic branches; but there is only the one before the measurement. If we suppose, contrary to the preceding account, that persons are localized in time, and are fully described by a quantum state (like the initial state $|0\rangle$), then prior to the measurement it seems there can be no self-locating uncertainty.[11] But look again. Eq. (3) can equally be written:

$$c_+ |\phi_+\rangle \otimes |0\rangle + c_- |\phi_-\rangle \otimes |0\rangle \xrightarrow[U_M]{} c_+ |\phi'_+\rangle \otimes |+\rangle + c_- |\phi'_-\rangle \otimes |-\rangle. \qquad (4)$$

There is an observer in the state $|0\rangle$ correlated with the state $|\phi_+\rangle$ (but who does not know it), prior to the measurement interaction, who will unitarily evolve to record the value plus, and there is an observer correlated with $|\phi_-\rangle$ (who does not know it), similarly without interaction, who will go on to record the value minus. Why does Eq. (3) suggest the right reading, and not Eq. (4)?

An added principle seems to be needed to rule out reading Eq. (4) in this way, perhaps a version of Leibniz's principle of identity of indiscernibles, locally enforceable. For those so insistent, still an account of uncertainty can be given, shifting the period of uncertainty from immediately prior to the measurement, to a time immediately following. The observer, we may suppose, simply *closes her eyes*.[12] Given which, post-measurement there are undoubtedly two observers present, with different properties (denote $|0_+\rangle$, $|0_-\rangle$), neither as yet aware of the outcome. In place of Eq. (4), we have:

$$|\psi\rangle \otimes |0\rangle \xrightarrow[U_M]{} c_+ |\phi_+\rangle \otimes |0_+\rangle + c_- |\phi_-\rangle \otimes |0_-\rangle \qquad (5a)$$

$$\xrightarrow[U_O]{} c_+ |\phi_+\rangle \otimes |+\rangle + c_- |\phi_-\rangle \otimes |-\rangle \qquad (5b)$$

where $U_M$ is as before the measurement process, yielding a superposition of macroscopically distinct outcomes, and $U_O$ represents the further act of observation by the observer of what that outcome is. After Eq. (5a) but before Eq. (5b), there are unequivocally distinct observers, for the two states $|0_\pm\rangle$ ex hypothesis differ physically, each ignorant of which state they are in, and with which state they are correlated. Eq. (4) gives momentary pre-measurement uncertainty, and Eq. (5a) gives momentary post-measurement pre-observation uncertainty. Arguably, anticipating uncertainty to come is a form of pre-measurement uncertainty.[13]

A third and still weaker notion of uncertainty appeals only to behavior.[14] Call *predictive behavior with respect to E* behavior or action that (i) predicts $E$ and (ii) commits solely to $E$, which selectively anticipates $E$ to the exclusion of all other possibilities. Much of our behavior is predictive, in social interactions as in the natural world, as night follows day. But for other events, among them ruinous events, predictive behavior is often impossible – not because we cannot prepare for them but because we do not know when or whether they will occur. We cannot commit to more than one, by definition of "commitment"; we can of course still commit to just one, so at least in that eventuality we are fully prepared – and be right in that eventually, but then we will be wrong otherwise. Better, we do not commit to one eventuality at all; we prepare for all or several of them, taking out an insurance policy instead. We do not behave predictively; we entertain several possibilities; we call this uncertainty.

Consider now branching in quantum mechanics, in the special case where the branching structure is fully known, as in performing a simple quantum experiment. Let some of the ruinous events be among the outcomes of the experiment (it is almost as bad as Schrödinger's cat). Still predictive behavior is impossible, not because events $E$, $E'$, $E''$ ... on different branches are unpredictable, as happening at unforeseen times, but because they are all happening at the same foreseen time. Any action initiated prior to measurement will take the same form in every branch, and if it is especially fitted to one, it will be unfitted to all the others. We can still commit to just one, so that at least in that branch we will be fully prepared – but at the price that in all the others we are unready.

Resourcing events that occur unpredictably, at different times, is no different from resourcing events that occur at the same time, in different branches. Predictive behavior in both cases is impossible. A limited transfer of resources, from the times and branches in which ruinous events do not intrude to the times and branches in which they do, is the rational strategy. We should *behave* as we do when we are uncertain. Uncertainty in the face of branching, on this approach, does not require lack of propositional knowledge, or lack of self-locating knowledge, but reflects rather the lack of an ability: knowing everything there is to know, we still cannot act predictively. It is because the same ability is lacking when predictions cannot be made, our usual predicament, that we think probability must involve ignorance. Degrees of belief are important because they are salient to action; so too, and more directly, are degrees of commitment.

## 2. The Connection with Statistics

A second standard objection to the identification of probability with branch weight is that the connection with statistics is wrong. To take the measurement of spin with protocol (1), after $N$ repetitions, with the same initial state (2) prepared for each trial, the result is a superposition of $2^N$ branches, one for each possible sequence of outcomes, of varying weight depending on the sequence. Everett's insight was that for large $N$, the collective weight – the squared amplitude – of the superposition of all those branches with anomalous statistics falls off exponentially in $N$ in comparison to the weight of all those with the correct statistics.[15] This is an example of a quantum law of large numbers (a "quantum Bernoulli theorem"). It is essential for the interpretation of branch weights as probabilities, but it is not sufficient for it; it does not, for example, imply that anomalous histories are not there – only that they have low collective weight. For those who understood the theorem as an attempt to *derive* probabilities from statistics, the attempt fails.[16] That was not, however, Everett's intent, which was rather to show parity with the way probability enters in classical statistical mechanics.

But is there parity? It is true that on any theory of probability there is a non-zero probability of anomalous statistics; but in a one-world setting, events of small enough probability may never happen. However, they *will* happen, eventually, in a one-world stochastic theory for sufficiently many trials, and likewise in a one-world deterministic theory for a world of finite volume, given sufficient time. In a spatially large enough world, deterministic or stochastic, they happen, somewhere, all the time. Given the size of the visible universe, and in the sure knowledge that it extends far beyond our event horizon, it is hard to set much store on the argument that the key difference with probability in Everett's approach is that in one-world theories the universe *may* in fact be small enough not to contain anomalous statistics. [17]

Denizens of anomalous branches, or of anomalous stretches of history in one-world theories, will be misled by the observed statistics of measurements. They will conclude that quantum mechanics (or at least the Born rule) is false. But they will simply be unlucky. We already have to live with this: the earth is large enough, and people are numerous enough, to play out the same argument. There are those among us who have been struck by lightning many times: how can they not believe, in their heart of hearts, that someone is out to get them? – and that someone is protecting them too. They are epistemically unlucky. We are used to this.

Questions of epistemology go over to questions of measurability. How can branch weights be measured? In the Everett interpretation, this is a purely *dynamical* question. It is the question of how amplitudes as they figure in a state of the form (2) can be reliably correlated with macroscopic indicators (a "probability meter"). The answer is that only the *ratio* of the amplitudes can reliably be measured, where "reliable" means: in branches with collective amplitude close to unity, when the number of trials is large. We are back to the law of large numbers, derived from the unitary dynamics of quantum theory. In the same way, the Kolmogorov axioms themselves are derived as approximate, "high-level" laws. We are used to abstract theories of geometry, as opposed to physical geometry; there is likewise abstract probability theory, as opposed to physical probability.

In one-world chance theories, how chances are measured will depend on what those theories say those chances are. This is work in progress (what, exactly, are chances in classical statistical mechanics, or in pilot-wave theory?), but no one expects to do any better than our actual practice. They will say: it follows from the *concept* of probability, that the best we can do is to measure relative frequencies, which will probably match the probabilities. However, if chances are the squared amplitudes of macroscopic branches, the Everett interpretation explains *why* they can only be measured in this way, as follow from dynamical considerations.[18]

There are even those who criticize the Everett interpretation because of these limitative results; who insist that there must be a causal mechanism that will reliably bring about true beliefs about the amplitudes, where "reliable" does not involve probability; that it must be possible, if the unitary evolution is all that there is and if the theory is to be empirically adequate, for ratios in branch weights to be deterministically driven in to the memory of some measurement device

– and if this cannot be ensured then the Everett interpretation must be rejected.[19] But no conventional theory of probability delivers so much; why demand it of quantum mechanics under the Everett interpretation?

## 3. Decoherence Theory

Two other objections will occupy us in this section and the next, both specific to the Everett interpretation. They both concern the reductive project itself, of giving an account of what probabilities are in terms of something that at first sight is not chancy. But some more stage-setting is needed.

Everett offered up the picture of a branching structure to the wave-function in which branching was defined by measurement interactions; he had no account of it otherwise. But as a realist interpretation of quantum theory, branching cannot arise *only* with quantum measurements, as if, fantastically, only a single world existed before quantum mechanics was discovered, and before any quantum experiments were performed. How, in the absence of measurement interactions, does branching arise, and with respect to what basis?

This was called the "preferred basis problem."[20] The solution lay in decoherence theory – roughly speaking, the theory of how the components of superpositions are subject to "effective" equations, yielding approximately classical behavior for the components, as a consequence of the unitary dynamics by which superpositions propagate as wholes. (for more background see the companion paper.) Where the rules break down (because only approximately satisfied), or where basis states propagate in the way of equations with dissipation and noise, each basis state evolves into further superpositions of basis states. Branching structure made out in this way is *emergent*: it involves approximations and the identification of salient dynamical variables, in much the same way that emergence is made out across the special sciences. But branching just is chancing; hence so too is chance. Physical probability is something emergent, along with classicality itself.[21]

All this structure, thus revealed, is needed to show that branch weights play the chance roles – to obtain branching and branch weights to begin with. But decoherence theory itself involves probability. Take, for example, Ehrenfest's theorem in the case of an initial state $|\psi\rangle$, well-localized in position and momentum, important to Everett's argument for branching (see the companion paper). There we define a quantity $\langle\hat{x}\rangle_\psi$ that for sufficiently well-behaved potentials can be shown to satisfy classical equations. But the quantity $\langle\hat{x}\rangle_\psi$ has a probabilistic interpretation: it is the expectation value of the position operator $\hat{x}$ in the state $|\psi\rangle$, and it is measured by repeated experiments, invoking the Born rule.

For another example, take the concept of a quasiclassical domain, as first defined by Murray Gell-Mann and James Hartle in 1989, likewise important to the Everett interpretation (see the companion paper). It is a history space "with probabilities peaked on quasiclassical histories." According to Adrian Kent, a prominent critic, this shows that "the ontology is *defined* by applying the Born rule" (Kent 2010, pp. 337–338). Jonathan Halliwell, who has widely applied decoherent histories theory to quantum foundations, likewise speaks in probabilistic terms. For example, for a quasiclassical domain defined by hydrodynamical variables, he writes:

> The final picture we have is as follows. We can imagine an initial state for the system which contains superpositions of macroscopically very distinct states. Decoherence of histories indicates that these states may be treated separately and we thus obtain a set of trajectories which may be regarded as exclusive alternatives each occurring with some probability. Those probabilities are peaked about the average values of the local densities. We have argued that each local density eigenstate may then tend to local equilibrium, and a set of hydrodynamic equations for the average values of the local densities then follows. We thus obtain a statistical ensemble of trajectories, each of which obeys hydrodynamic equations. These equations could be very different from one trajectory to the next, having, for example, significantly different values of temperature. In the most general case they could even be in different phases, for example one a gas, one a liquid. (Halliwell, 2010, p. 111)

The criticism, after all this stage-setting, is this. Probability talk is ubiquitous in the literature on decoherence theory. In order to have meaning, probabilities have to be assumed from the outset. But then it follows that there is no reductive, Humean base level of description, free of probabilistic reasoning, on which probabilities supervene. When it comes to probability in Everettian quantum mechanics, the project of Humean supervenience cannot even get off the ground.

The point at issue, however, is not whether models of decoherence theory as *usually* derived and discussed involve probability; we grant that they do. Nor is it a surprise that these probabilities are interpreted in one-world terms (as in Halliwell's writings): from its inception the decoherent histories theory was supposed to provide a one-world interpretation of quantum theory, without any need of Everett's extreme ideas. The substantial objection can only be that these models *cannot be divested* of their probability interpretation, and of probability theory, not even on going over to the Everett interpretation.

Is this true? To take the case of Gell-Mann and Hartle's definition of a quasiclassical domain, here is a replacement formulation: it is a space of histories for which the amplitudes are strongly peaked on histories obeying a closed system of equations. "Strongly peaked amplitude" does not, prior to defining the branching structure of the state, have to be interpreted as "highly probable." Halliwell's summary can similarly be reworded, noting that the "average values of local densities" are defined not by averaging the densities, but as the values of the local densities on those trajectories on which the amplitudes are (very sharply) peaked. In the case of Ehrenfest's theorem, whilst it is *possible* to interpret $\langle \hat{x} \rangle_\psi$ operationally, in terms of multiple measurements (assuming similar systems can be prepared in the same state $|\psi\rangle$), it is also possible (when $|\psi\rangle$ is sharply peaked in position and momentum) to interpret it realistically, as the location of the peak of the wave-function as it evolves over time, in accordance with classical equations (as in the companion paper).

To give another example, take the requirement of consistency among histories (the condition that the so-called "sum rules" be satisfied). This is supposedly an a priori constraint on any probability theory on a history space. In the quantum histories formalism, it forces the vanishing of the real part of the inner product of states of distinct histories, but that in turn can be directly related to interference between histories, which is hardly built into the concept of probability. In fact the stronger condition, that real *and* imaginary parts vanish, is both more natural and far more widely used. As such it is the requirement that the structure to the quantum state, as defined in terms of quantum histories, is to be made out in terms of orthogonal vectors: in terms of a basis. Orthogonality is as useful to get at the structure of the state over a period of time, as it is for the structure to the state at an instant of time. Probability need not come into it.[22]

The objection may concern justification rather than understanding, particularly if decoherence theory is used to derive the Born rule. According to Wojciech Zurek, an early pioneer of decoherence theory, concepts like "partial trace" and "reduced density matrix" cannot be used for that purpose "because their physical significance depends on Born's rule."[23] For a more recent critique:

> In order to neglect small values in favour of larger values, we have to establish that the magnitude of the corresponding variable is related to the entry's effect on the measurement to be performed. Since experimental testing and the entries in the density matrix are related in terms of the probabilities for measuring certain outcomes, in order to establish the negligibility of small entries in the density matrix we must introduce the Born rule.

There is another way to analyze the effect on the measurement to be performed, however: model that measurement device explicitly in the formalism – precisely Everett's method. So long as we may interpret the quantum state that results, for example, in terms of a macroscopic pointer position, it can be established whether it depends on those small entries – on whether the off-diagonal elements in the reduced density matrix (for example) can be neglected, without consequence. And pointer positions, as we have just seen, move about in three-dimensional space, within the limits of Ehrenfest's theorem, as do classical particles, and can be picked out on that basis without any need for probabilities – despite the fact (this being the Everett interpretation) they occur in superpositions.

As W.V. Quine said, a physical theory is tested as a whole; it is the exception when different parts of it can be isolated as independently testable. It is a virtue where it is possible, not a requirement to which theories must be held accountable. This is "meaning holism", otherwise known as "the Duhem-Quine thesis", and it has been widely accepted in philosophy of science in the last century.

## 4. Branch Counting

Given the branching structure to the universal wave function, it is clear what is the intended interpretation of probability (ratios in branch weights). Are we sure there is no rival alternative? There is one that has been taken seriously even by those sympathetic to Everett's ideas: the *branch-counting rule*.[24] It is the rule that on any branching event, all outcomes, all histories that have ensued at any given time are equiprobable. If from repeated measurements a large slew of histories result, the number with a given relative frequency (divided by the total number) determines the probability of that relative frequency.

The result, for the Everett interpretation, is mayhem. To take again the measurement of spin with initial state (2), after $N$ measurements the vast majority of states have relative frequency of plus-outcomes equal to one half, and likewise for minus-outcomes one half, *entirely independent of $|c_+|^2$ and $|c_-|^2$*. When this ratio differs significantly from unity, only a *tiny minority* of branches after $N$ trials comply with the Born-rule.

The branch-counting rule makes nonsense of quantum mechanics but it appears to be suggested by the picture of branching. It spells trouble for a reductive approach to probability if the base provides two quite different candidates for the chance role, ratios in branch weights, and ratios in branch numbers. We know which one is wrong, on empirical grounds, but what makes the one *probability*, and not the other? Probability, it suddenly seems, may have to be taken as a primitive after all.

In point of fact (as shown by Wallace 2012), this probability rule – call it "naive" branch counting – conflicts rather straightforwardly with axioms of probability. Thus, consider Eq. (5a), and suppose, in place of Eq. (5b), a second measurement is made at time $t_1$, but only in the branch with the plus-outcome, of something else entirely, say position, producing two further branches at $t_2$, each with the plus-outcome. In the branch with a minus-outcome, no further measurement is made, so there is only one such branch at $t_2$. So what is the probability of the plus outcome at $t_2$? At $t_1$ it was one-half, at $t_2$ two-thirds, but the latter cannot be obtained by updating in time, for it is in conflict with the sum rule:

$$\Pr(+; t_2) = \Pr(+; t_2/+; t_1)\Pr(+; t_1) + \Pr(+; t_2/-; t_1)\Pr(-; t_1) \qquad (6)$$

which follows from the probability calculus for histories when the probabilities of plus and minus outcomes at $t_1$ sum to unity. Using naive branch counting, Eq. (6) yields one-half, not two-thirds. [25]

It follows that the naive branch counting rule is not, in fact, a coherent probability rule at all. If this were the only alternative to the Born rule, there would be no problem of underdetermination as alleged. But there is another branch-counting rule that is if anything more intuitive: it is that on each branching event, the probabilities of each branch thus produced are the same, with probabilities for branches at subsequent times not equiprobable, but depending on how each branch came about, all conforming to sum rules of the form (6).[26]

This new branch-counting rule is in general just as hopelessly at odds with the Born rule; moreover, it is manifestly in conflict with locality (as permitting super-luminary signaling). [27] But that is only grist to the skeptic's mill: the skeptic is arguing that the branching structure to the state on measurement, with number equal to the number of possible outcomes, the central concept of the Everett interpretation, suggests an altogether inappropriate concept of probability. If it is at odds with relativity as well as the Born rule, so much the worse for the Everett interpretation.

But is this branching number so defined central to the Everett interpretation? It was prior to decoherence theory, when appeal to experiments, with a definite number of possible outcomes, was the only way that branching was defined. But decoherence theory changed all that. Decoherence theory, as argued in the companion paper, just is the theory of branching structure, but as it occurs naturally, independent of whether any experiments are performed.

Using decoherence theory, when measurements are made, branching number has nothing to do with the number of *readings* that can be made. Consider again the measurement of spin, and specifically, consider just one of the two protocols, say Eq. (1a). Is there just one way this experiment can come about – as a physical process? Consider all the quasiclassical processes going on – the thermal fluctuations, variations in pressure, Brownian motions, cascades of phonons, scattering of light – all of them involving branching, and all that *just on opening the laboratory door*. There are clearly *countless* different ways that the apparatus can obey Eq. (1a), evolving from macrostates in which it reads "ready" to those in which it reads "+" even when the initial spin state that is measured is always $|\phi_+\rangle$. Here, "countless" means undefined: the number of branches, specified by what goes on in each branch during the process of measurement, is undefined.

This could be work in progress. It may be, for example, that there is a finest-grained history space that is a decohering history space, indeed, a quasiclassical domain – and we just don't happen to know how to approximate it. We use a convenient definition, and relative to this the branch number is fixed; the number is somewhat arbitrary to be sure, depending on the coarse-graining, but it might be thought of as our best guess on what the "correct" fine-graining is. If it can be shown that the probabilities thus defined (as ratios in finite numbers) are insensitive to this coarse-graining, including coarse-graining in time, the new branch-counting rule would be definable after all – to the possible discredit of the Everett interpretation.[28]

The true nature of the difficulty only now comes into focus. For of all of these branches, thus defined, waiting to be counted: is it all of them, or only the ones with non-zero amplitudes? Neither choice makes any sense. If it is all of them, then the numbers are completely independent of the state; in what sense is this an interpretation of the structure of the state? But if only non-zero amplitude branches are counted, the numbers will be discontinuous functions of the amplitudes. The tiniest variation in amplitude may make for arbitrarily large variations in branch numbers.[29] Yet continuity in the amplitudes (in the Hilbert-space norm, the norm topology) is essential to decoherence theory and the unitary dynamics; as likewise to spectral theory, and the whole edifice of quantum mechanics.

The two objections to Everett's approach – that decoherence theory must be founded on a probability rule, and that one such rule is the branch-counting rule – can now be nicely brought together. Given the branch-counting rule as just stated, indeed no approximation can be made, no low-amplitude history neglected, without checking with that rule. But it is a Pyrrhic victory, twice over. Branch counting is hardly a rival to the branch-weight rule as an interpretation of branching structure, if it brings that branching structure crashing down. And of course the tiniest of approximations used in a theoretical model must be justified by the probability rule, if the latter is so patently at variance with the mathematical structure of the theory.

The supervenience project goes precisely the other way. Branch numbers can always be defined in terms of the branch weights, by the requirement that they have equal weight. The absolute numbers thus arrived at are arbitrary, to be sure, but *ratios* in those numbers are well-defined, free of any convention, in agreement with the Born rule.[30]

## 5. Undermining

The philosophical literature on reductive "Humean" theories of chance has led to a certain consensus. Such a theory should satisfy the following:[31]

(i) The Principal Principle – one should set one's credence in $E$ at $t$ equal to the chance of $E$ at $t$, no matter what else one knows, provided one has no magical (no "inadmissible") information from the future.

(ii) Quantitative constraints – the chance of an event after it occurs is always 1 or 0; an event that has value 0 or 1 at one time retains that value for all subsequent times.

(iii) The chance-frequency link – the relative frequency of $E$s in an ensemble of systems all prepared in the same state approaches the probability of $E$ in that state as the size of the ensemble increases, but the *possibility* of divergence in any finite ensemble remains, no matter how large the ensemble.

It has led to a consensus, in particular, that in a conventional Humean setting (meaning, inter alia, a one-world setting) no such theory has been found. Indeed, many conclude no such theory *can* be found, that there can be no "perfect" theory of chance.[32] The difficulty is that the link between chance and the reductive base must be slack – to countenance (iii) – so, for example, there must be possible worlds in which the statistics are the same and the chances different, or the chances the same and the statistics different; but no, the link has to be tight, the chances cannot drift far from actual events. If the two were distinct existences, there would be a world where $E$ occurs at time $t$, but where the chance of $E$ at some later time is not one, contrary to (ii).

Add to these the Basic Chance Principle (BCP) (Bigelow et al., 1993), the principle, roughly, that the chance for an event $E$ at some future time may be the same, as determined by events up to some earlier time, whether or not $E$ actually happens. Indexing to worlds and to times it is the principle:

(iv) BCP – if the chance of $E$ at world $w$ at time $t$ is $P_{tw}(E) > 0$, then there exists a world $w'$ which (a) matches $w$ up to time $t$, (b) contains $E$, and (c) satisfies $P_{tw}(E) = P_{tw'}(E)$.

If there were no such world $w'$, then the chance of $E$ at $t$ could not be the same independent of whether or not $E$ happens – for if the BCP fails, there is no world containing $E$ for which the chance at $t$ is the same. Yet it seems the BCP must fail – either that, or there are no patterns of events $E, E'$, that yield distinct chance theories in any worlds $w, w'$ in which $E$ and $E'$ occur; distinct, in particular, in that at some $t$, $P_{tw}(E) \neq P_{tw'}(E)$, and $P_{tw}(E') \neq P_{tw'}(E')$.

This is an instance of "undermining." A similar argument shows that (i), the Principal Principle, must fail; in effect, it shows that knowing the chance at $t$ for such patterns *is* to have magical information about the future – is itself inadmissible – because what that chance is depends on what happens at future times.[33] But if there are no patterns of this kind, the very idea of a Humean reductive theory of chance is in trouble.

*Back to Everett.* A quasiclassical history, divested of amplitude and phase, just is one local pattern of events after another; each is a Humean tapestry of events, a Lewisian world $w$. The Everett interpretation thus provides a home for Lewis's metaphysics, but with these essential differences: the worlds are emergent structures, so no best-system analysis based on them can hope to give the fundamental laws (at most they may give the emergent laws); the worlds bear new and irreducible relations to each other, defined in terms of amplitude and phase; and the worlds have branching structure,[34] as defined by these amplitudes and phases.

On the most straightforward identifications, Lewisian worlds correspond to quasiclassical histories,[35] but the reductive base is the collection of all these histories in a superposition – a branching structure – or, in Lewisian terms, a collection of worlds. The chance theory for this collection, arranged in (derived from) this branching structure, is that chance events are branching events, and chances are ratios in branch weights: chancing is branching. By that theory, for any world $w$ at time $t$, the chances are determined only by its history up to time $t$, for that alone (with the unitary dynamics) determines the amplitudes of branching events thereafter. They are the same (as functions $P_{tw}$ from an event space to chances) for all worlds $w$ in the branching structure that match up to time $t$. There is just the one set of chances, at a branching event, regardless of subsequent branching events.

This theory of chance meets the principles (ii)–(iv) to perfection. The indexing to times in (ii) is automatically taken care of: chances are relations between branch weights, indexed to times, and they are retrodictively 0s and 1s because of branching structure (no recombination of branches; see the companion paper). (iii) is satisfied, reading "possibly" to mean there exist anomalous histories, albeit of vanishingly small amplitude. (iv) is satisfied: for *every* world $w'$ that matches $w$ up to time $t$, the chances at $t$ of $E$ are the same, for chances at $t$ are determined by the prior history at $t$. There is no undermining: the conclusion of the undermining argument was that either the BCP is wrong or (roughly) present chances in world $w$ are not determined by future events in $w$; but the latter is now a feature of the chance theory, not a bug.

How is it a feature, and isn't there a cost? Yes: what was before an instability in what the chances really are (if dependent on future chance events, as well as past), now shows up as a mismatch of relative frequencies to branch weights in some branches. It has turned into the problem of anomalous statistics, already considered. There are worlds whose inhabitants will be misled by the observed statistics, the epistemically unlucky ones. In the overwhelming majority of worlds,[37] their inhabitants will be led to the right theory. The Born rule remains the best-systems analysis of all the branching worlds, but it does not supervene on the statistics in each world, rather it supervenes on – is a simple function of – the branch weights of all the worlds. Could the denizens of the unlucky worlds survey the entire branching structure, their favored best systems analysis would be the Born rule.

There remains (i), Lewis's famous Principal Principle. But here too there has been progress.

### 6. The Principal Principle

Failing an account of what chances are, as indexed to times, it was obscure why credence as indexed to times should conform to them.[38] It was already obscure why they should if chances are relative frequencies of events up to that time, although there at least the obscurity was dignified as a philosophical problem (the "problem of induction"). In this context, the recently proved Born-rule theorem is little short of a philosophical sensation: it is the derivation of the Principle Principal in the special case where the chances are identified as branch weights and chance processes are identified as branching processes. Alternatively, assuming the Princpal Princple, it shows that branch weights *should* be identified as physical probabilities, for the latter are whatever satisfy the Principal Principle. Either way, it shows why credences should conform to branch weights.

The result turns on the principle of indifference already announced, but it also depends on rational choice theory, and, specifically, on the operational techniques first introduced by Frank Ramsey and Bruno de Finetti in the early part of the last century, whereby credences are operationalized in terms of betting behavior. The Dutch book argument is a case in point: agents were sure to lose money whatever the outcomes of bets, if their betting quotients did not conform to the axioms of probability theory. In Leonard Savage's 1954 landmark, *The Foundations of Statistics* (published by Princeton just as Everett began his studies), this took the form of a representation theorem: if the preferences of an agent among bets (of sufficient number and variety) conform to certain rules in decision theory, then there is an essentially unique credence function and utility function, such that those preferences are the same as by ranking by expected utility. That one's credence and utility function should dictate a rank ordering is obvious, it is the converse, perhaps, that is surprising. Still, that credence function, other than qualifying as a bone fide probability distribution, could be anything, as likewise one's utilities.

Fairly obviously, it will be impossible to tie down the credence function further without knowing more about how the games are actually played, and what the agent knows about how the games are actually played – in short, without a physical theory governing those games, and agents who base their choices among games on that theory. But as before it is essential, if it is to serve its purpose, that that theory be divested of any probability interpretation: wave mechanics without probability to the rescue again. Moreover, it would be better if that theory is divested, even, of any talk of uncertainty, since that notion too is contested.[39] Given all of which, the remarkable result first proved by Deutsch and as strengthened by Wallace is that wave mechanics, thus disinterpreted, is enough to tie down that credence function uniquely, to the point that it must conform to the branch weights. Agents, if rational, basing their choices on unitary quantum mechanics, and fully cognizant of the branching produced by each quantum game (as determined by the physics of the apparatus) and knowing the stake for each game and the rewards on each outcome, will order their preferences among games as if maximizing their expected utilities, for some utility function, using the Born rule. Difference in utilities will make a difference to their preferences, but their credence functions will always be the same: the Born rule.

The core of the proof is a symmetry argument.[40] Consider, for the last time, the measurement of the *z*-component of spin. Suppose an agent is to bet on the spin-plus outcome of the experiment, where the measurement process satisfies Eq. (1), that is, according to the protocol:

$$P1: \quad |\phi_+\rangle \otimes |0\rangle \xrightarrow[U_1]{} |+\rangle$$
$$|\phi_-\rangle \otimes |0\rangle \xrightarrow[U_1]{} |-\rangle$$

(where we assume that the agent doesn't care what happens to the state of the measured system

$|\phi_\pm\rangle$ after the measurement, so it is just omitted). Let the agent's credence in a spin-up outcome conditional on this protocol be $Cr(+/P1)$. Suppose now the protocol is changed to $P2$, according to which the same experiment is run, save that after the measurement the plus outcome is replaced by the minus-outcome, and vice versa. So the new protocol is:

$$P2: \quad |\phi_+\rangle \otimes |0\rangle \xrightarrow[U_1]{} |+\rangle \xrightarrow[U_2]{} |-\rangle$$

$$|\phi_-\rangle \otimes |0\rangle \xrightarrow[U_1]{} |-\rangle \xrightarrow[U_2]{} |+\rangle.$$

Then $Cr(+/P1)$ should be equal to $Cr(-/P2)$, since $U_2$ only swaps the outcomes after the measurement has been performed. But it follows that when the initial state is $|\psi\rangle$ as given by Eq. (2), the final state on $P1$ is (cf. Eq. (3)):

$$|\psi\rangle \otimes |0\rangle \xrightarrow[U_1]{} c_1|+\rangle + c_2|-\rangle$$

whereas on $P2$ it is:

$$|\psi\rangle \otimes |0\rangle \xrightarrow[U_2 U_1]{} c_1|-\rangle + c_2|+\rangle$$

and in the particular case when $c_1 = c_2$ *the two states at the end of the measurement are exactly the same, whichever protocol is used*. Therefore, the agent should be indifferent (for this special case) which protocol is used. So:

$$Cr(\pm/P1) = Cr(\pm/P2). \tag{7}$$

We already have:

$$Cr(\pm/P2) = Cr(\mp/P1).$$

So in the special case $c_1 = c_2$, it follows from Eq. (7):

$$Cr(\pm/P1) = Cr(\mp/P1)$$

and likewise for the protocol $P2$. The agent's credences for the two outcomes, on either protocol, should be the same.

Evidently, the argument at crucial moments appealed to normative judgments. The two states at the end of the two experiments, for the two protocols, won't be *exactly* alike; but the agent shouldn't care about microscopic inessentials. In extending the argument to rational ratios of $|\alpha|$ and $|\beta|$, ancillary devices are needed, which produce additional branching; but the agent shouldn't care about that either, if the branches produced differ only in ways the agent doesn't care about. The agent should only care about what is physically realized in each state (what can be given a dollar value in each state), and so on.

Many of these normative judgments are based on pragmatic constraints, on the basis of that old adage, "ought" implies "can"; so contraposing, you ought not to care about branching per se, because you can't (branching as determined by decoherence theory is ubiquitous). Others are more purely normative. The end result is a demonstration of why agents should care about branch weights, and why their credences should conform to them – and not, for example, to the number of four-leaf clovers on each branch, or the number of calibrations on the pointer dial.

Yet despite all these successes, the Born-rule theorem, for those convinced that the notion of probability in the Everett interpretation is otherwise unintelligible, has been found wanting. For them the theorem must carry the entire burden of probabilistic reasoning – whereby the explanation for the statistics of quantum experiments, normally provided by the Born rule, has to be provided instead by the betting strategies of experimentalists. But how can someone's betting strategy explain why the radium atom has a half-life of less than a second? How can the Born rule, important to physics, be true by virtue of human behavior? Doesn't it follow that there is no such thing as probability in a universe without people? Moreover, are there not alternatives to using the branch weights, no matter if fanciful or practically impossible to implement, that may not yet be irrational?[41] But all this is to take the Born-rule theorem in isolation from the larger reductive project.

The theorem demonstrates that the particular role ordinarily but mysteriously played by physical probabilities, whatever they are, in our rational lives, is played in a wholly perspicuous and unmysterious way by ratios in branch weights and branching, when known. It is because these quantities play all the other chance roles as well, that they deserve to be called probabilities.

---

[2] This is a slightly revised version of Chapter 15 of the same name in *The Routledge Companion to Philosophy of Physics*, E. Knox and A. Wilson (eds.).

[2] This is the position of Deutsch (1999, 2016) and Greaves (2004) (under the rubric "the fission program"), and of Vaidman (in numerous publications; see Vaidman 2021). It has also been defended by Brown (2013) and Brown and Porath (2020).

[3] It eventually appeared as "The theory of the universal wave function" in DeWitt and Graham (1973).

[4] For more in this vein see Papineau (2010).

[5] See e.g. van Fraassen (1980), Ch.6.

[6] Although the concept of equiprobability played a fundamental role in the discovery of quantum mechanics. For a recent review, see Saunders (2020).

[7] The claim is sweeping but appears to be accurate. See Tipler (2014), Brown and Timpson (2016).

[8] A related worry is that quantum mechanics under the Everett interpretation may not be testable. For arguments that it is, see Greaves (2004), Greaves and Myrvold (2010) and, for a unified treatment including the Born-rule theorem, Wallace (2012, Ch.6). See also Deutsch (2016). If the argument of the text succeeds, these worries are greatly ameliorated.

[9] Saunders (2021), to appear as Chapter 14 in *The Routledge Companion to Philosophy of Physics*, E. Knox and A. Wilson (eds.).

[10] See Saunders and Wallace (2008a, 2008b), Saunders (2010) and, for a contrary view, Tappenden (2008). For further discussion, see Wallace (2012, Ch. 7) and for a comprehensive metaphysics, Wilson (2020).

[11] As insisted e.g. by Kent (2010, pp. 346–347), and Vaidman (2021).

[12] So-called "Vaidman uncertainty," first introduced in Vaidman (1998).

[13] Tappenden (2011). See also Sebens and Carroll (2016), McQueen and Vaidman (2019) for recent derivations of the Born rule framed in terms of Vaidman ignorance, together with a locality assumption.

[14] For the case of linguistic behavior, see Saunders (1998). The argument that follows is more in the spirit of Ismael (2017).

[15] See e.g. Saunders (2010), Wallace (2012 p.140).

[16] As for example Smolin (2019, p. 151).

[17] See Tegmark (2010) for more in this vein.

[18] Or whether indeed some novel method of measurement might yet be found. I am not hopeful, for reasons given in Saunders (2010).

[19] This seems to be the position of Adlam (2014); and perhaps also Rae (2009).

[20] Often attributed to Ballentine (1973), although he does not use the term.

[21] See Wallace (2003) and, for probability as emergent, Saunders (2005). Branching as emergent structure is an important theme in Wallace (2012, Ch. 1–2).

[22] This suggests consistency is a very weak condition, and so it is, as was shown by Dowker and Kent (1996). In the companion paper, the theory of quasiclassical histories is presented without any probability interpretation.

[23] Zurek (2005, p. 1); the quotation that follows is from Dawid and Thébault (2015).

[24] As first discussed by Graham (1973). It was at center stage in David Lewis's only foray into quantum mechanics (Lewis 2004).

[25] The parallels with the "Sleeping Beauty" paradox are evident; for this see Elga (2000).

[26] See Lewis (2009). It is ruled out by Wallace's axioms (under the "fake-state" rule), but critically because it violates certain "richness" axioms (in particular problem and solution continuity; see 2012, pp. 170–171, 178, pp. 91–92), so ultimately, for reasons similar to those given in the text.

[27] As was recently shown by McQueen and Vaidman (2019).

[28] However, such a stability condition seems likely to force the branch weight rule instead: see Saunders (2005).

[29] While the *bare* number of branches of non-zero weight, for a sequence of states $|\phi_k\rangle$ that converges to $|\phi\rangle$ in the norm topology, may not converge to anything, the *weighted* number converges smoothly. Thus for the region $\Delta_+$ corresponding to the + outcome, partitioned into cells $\delta$, the weighted number in the state $|\phi_k\rangle$ is not $\sum_{\delta \subset \Delta_+}$, where the sum is over those $\delta$ for which $\langle \phi_k | P_\delta | \phi_k \rangle \neq 0$, but $\sum_{\delta \subset \Delta_+} \langle \phi_k | P_\delta | \phi_k \rangle = \langle \phi_k | P_{\Delta_+} | \phi_k \rangle$, which converges to $\langle \phi | P_{\Delta_+} | \phi \rangle$ as $|\phi_k\rangle \to |\phi\rangle$.

[30] Compare Boltzmann's combinatoric approach to probability, introducing a unit on phase space to provide a discrete count of states all of equal volume; ratios in those numbers are independent of the choice of unit if sufficiently small, and agree with ratios in Liouville measure (in the limit, the agreement is exact).

[31] The three criteria that follow are taken almost verbatim from Ismael (2009, pp. 421–422); the paragraph after is my attempt to paraphrase her subsequent argument.

[32] The terminology is due to Schaffer (2003) (to be precise, a "perfect" theory in his sense is a reductive theory that satisfies the PP and the BCP; see below).

[33] As Lewis eventually came to see (Lewis 1994), chances at *t* in a one-world best systems theory are themselves inadmissible at *t*. This points to a reconciliation of sorts between undermining and the PP (Lewis's "big, bad bug"; see Lewis (1986)).

[34] For the connection with overlap and divergence, see Saunders (2010) and Wilson (2011, 2020). Here, I use "branching" as in the physics literature, neutral as to overlap vs divergence.

[35] The entire branching structure – the collection of worlds closed under relations of amplitude and phase – also has a claim to be considered as a single Lewisian world. Or it might better be thought of as an "inner sphere" of worlds, in roughly Lewis's sense. See Wilson (2020).

[37] In the sense of note 30.

[38] For the best of recent attempts, see Hoefer (2007) and Schwarz (2014).

[39] Nor should they be compromised on later *interpreting* the axioms in terms of ignorance and uncertainty. (Of course it will do no harm if they are thus strengthened; see Wilson (2013) for an argument of this kind.)

[40] Here I closely follow Wallace (2012, pp. 151–152).

[41] For criticisms of this form, see Albert (2010), Kent (2010), and Price (2010), and for discussion, see Saunders et al. (2010, pp. 391–406).